\documentstyle[psfig]{mn}
\def\phn{\phantom{0}}
\def\phd{\phantom{.}}

\let\simgt\ga

\let\arcdeg\degr
\def\acknowledgements{\section*{Acknowledgments}}
\def\Sec#1{Section~\ref{sec:#1}}
\def\Fig#1{Fig.~\ref{fig:#1}}
\def\Tab#1{Table~\ref{tab:#1}}
\def\RV{\ifmmode R_{\rm V}\else\hbox{$R_{\rm V}$}\fi}
\def\RC{\ifmmode R_{\rm C}\else\hbox{$R_{\rm C}$}\fi}
\def\DV{\ifmmode\Delta\Phi_{\rm V}\else\hbox{$\Delta\Phi_{\rm V}$}\fi}
\def\DC{\ifmmode\Delta\Phi_{\rm C}\else\hbox{$\Delta\Phi_{\rm C}$}\fi}
\def\kms{\ifmmode{\rm km\,s^{-1}}\else\hbox{$\rm km\,s^{-1}$}\fi}
\def\G{G\,}
\def\zzpsc{\G29-38}  

\def\phnn{\phn\phn}
\def\l{\ifmmode\ell\else$\ell$\fi}
\def\m{\ifmmode m\else$m$\fi}
\def\n{\ifmmode n\else$n$\fi}
\def\altaffilmark#1{$^{#1}$}
\def\altaffiltext#1#2{\\$^{#1}${#2}}
\def\plotone#1{\centerline{\psfig{figure=#1,width=\hsize}}}
\def\nodata{$\cdots$}
\def\colhead#1{\multicolumn{1}{c}{#1}}

\def\nl{\\}
\def\tablecomments#1{\par\smallskip\noindent Notes. #1}
\def\phasedef#1#2{\ifnum#2>30\nodata\phn\phn\else\pl{#1}{#2}\fi}
\def\dnil#1#2{$#1\ifnum#2>0\pm\ifnum#2>9 #2\else\phn #2\fi
                 \else\phantom{0\pm0}\fi$\ignorespaces}
\def\done#1#2{$#1\ifnum#2>0\pm\ifnum#2>9 \dsplit#2\else 0.#2\fi
                 \else\phantom{0\pm.0}\fi$\ignorespaces}
\def\dtwo#1#2{$#1\ifnum#2>0\pm\ifnum#2>9 .#2\else .0#2\fi
                 \else\phantom{0\pm.0}\fi$\ignorespaces}
\def\dsplit#1#2{#1.#2}
\let\fr\done\let\al\dtwo\let\pl\dnil\let\av\done\let\pv\phasedef
\def\rv#1#2{\dnil{#1}{#2}\phd}\let\dv\phasedef
\let\rc\done\let\dc\dnil\let\rd\rv\let\dd\dnil
\def\tabsize{15.1cm}\def\apptabsize{15.1cm}
\def\addphn#1{\if2#1 #1\else\phn #1\fi}
\def\nsiz#1{\setbox0\hbox{0}\hbox to\wd0{\hss\rm #1\hss}}
\makeatletter
\def\ion#1#2{#1$\;${\small\rm\@Roman{#2}}\relax}
\def\@cite#1#2{#1\if@tempswa , #2\fi}
\let\cite\@internalcite
\makeatother

\begin{document}

\title{Surface motion in the pulsating DA white dwarf \zzpsc}

\author[M. H. van Kerkwijk, J. C. Clemens and Y. Wu]{%
        M. H. van Kerkwijk\altaffilmark{1,2,3},
        J. C. Clemens\altaffilmark{1,4}
	and
	Y. Wu\altaffilmark{5,6}
\altaffiltext{1}{Palomar Observatory, California Institute of
                 Technology 105-24, Pasadena, CA 91125, USA}
\altaffiltext{2}{Institute of Astronomy, University of Cambridge,
		 Madingley Rd, Cambridge, CB3 0HA, UK} 
\altaffiltext{3}{Astronomical Institute, Utrecht University, P.O. Box
                 80000, 3508 TA Utrecht, The Netherlands;
                 M.H.vanKerkwijk@astro.uu.nl}
\altaffiltext{4}{Department of Physics and Astronomy, University of
                 North Carolina, Chapel Hill, NC 27599-3255, USA;
                 clemens@physics.unc.edu}
\altaffiltext{5}{Theoretical Astrophysics, California Institute of 
		 Technology 130-33, Pasadena, CA 91125, USA}
\altaffiltext{6}{Astronomy Unit, School of Math.\ Sci.,
		 Queen Mary and Westfield College, Mile End Road,
		 London E1 4NS, UK; Y.Wu@qmw.ac.uk}
}
\maketitle

\begin{abstract} 
We present time-resolved spectrophotometry of the pulsating DA white
dwarf \zzpsc.  As in previous broad-band photometry, the light curve
shows the presence of a large number of periodicities.  Many of these
are combination frequencies, i.e., periodicities occurring at
frequencies that are sums or differences of frequencies of stronger,
real modes.  We identify at least six real modes, and at least five
combination frequencies.  We measure line-of-sight velocities for our
spectra and detect periodic variations at the frequencies of five of
the six real modes, with amplitudes of up to $5\,{\rm{}km\,s^{-1}}$.
We argue that these variations reflect the horizontal surface motion
associated with the g-mode pulsations.  No velocity signals are
detected at any of the combination frequencies, confirming that the
flux variations at these frequencies do not reflect physical
pulsation, but rather mixing of frequencies due to a non-linear
transformation in the outer layers of the star.  We discuss the
amplitude ratios and phase differences found for the velocity and
light variations, as well as those found for the real modes and their
combination frequencies, both in a model-independent way and in the
context of models based on the convective-driving mechanism.  In a
companion paper, we use the wavelength dependence of the amplitudes of
the modes to infer their spherical degree.
\end{abstract}


\section{Introduction}\label{sec:intr}


Pulsating white dwarfs show brightness variations at a range of
frequencies, with amplitudes of up to a few per cent.  They appear to
be normal white dwarfs, except for the fact that they lie in a special
temperature range, or instability strip, in which they are unstable to
pulsations.  There are instability strips for all three of the main
white dwarf types, DA, DB and DO (atmosphere dominated by \ion{H}{1},
\ion{He}{1} and \ion{He}{2}, respectively), around 12\,000, 23\,000
and 100\,000\,K, respectively.


Pulsating white dwarfs offer the prospect of physical insight via a
number of different methods.  One is asteroseismology, the
determination of a white dwarf's internal structure through the
comparison of observed periodicities with model predictions of stellar
eigen-modes.  Another method analyses which modes are present, what
their amplitudes are and how the amplitudes vary.  The aim is to learn
about the physics of the upper layers of the star, by determining the
way in which modes interact with different parts of the star and with
each other.


Robinson, Kepler \& Nather (\cite{robikn:82}) showed that white dwarfs
pulsate in non-radial gravity modes, and that the observed brightness
variations result from perturbations of their temperature, not of
their geometry or gravity.  They also estimated the expected
variations in the line-of-sight velocity and the line profile, and
concluded that these would be too small to measure with then available
instrumentation.  As we will show below, this statement no longer
holds.


The pulsations were originally thought to be excited in an ionisation
zone, as in classical pulsators such as the Cepheids.  The dynamics in
DA and DB pulsating white dwarfs are rather unlike those in classical
pulsators, however, in that for the white dwarfs the surface
convection zone can respond almost instantaneously to gravity-mode
pulsations, because the turnover times are much shorter than the mode
periods.  In a series of papers, Brickhill (\cite{bric:83},
\cite{bric:90}, \cite{bric:91a},\cite{bric:91b}) has described the
response in detail, and concludes it leads to significant ``convective
driving'', which will be sufficient to overcome the damping in the
radiative interior for modes with periods of order of or shorter than
the thermal response time of the convection zone.  Brickhill's partly
numerical results have recently been confirmed analytically (Wu
\cite{wu:97}; Goldreich \& Wu \cite{goldw:99a}).  


The DA and DB pulsating white dwarfs often show power at frequencies
that are the sums or differences of frequencies with stronger power.
Based on the density of the theoretical gravity-mode spectra in the
relevant frequency range, it has been argued that these ``combination
frequencies'' do not represent real eigen-modes, but are instead
produced by nonlinear mixing of pulsation signals of real eigen-modes
(``parent modes'') in the outer layer of the star.  Some support for
this idea comes from observed correlations in power.

Brickhill (\cite{bric:92a},b) has suggested that the mixing is a
by-product of the interaction between the pulsations and the
convection zone.  Because of the relatively slow thermal response of
the convection zone, flux variations take some time to appear at the
surface, and they are reduced in amplitude.  But the thickness of the
convection zone is modulated by the pulsation, so that the signal is
not only diminished and delayed, but also distorted.  It is the latter
distortion which translates into the power seen at combination
frequencies.

Both to be able to make an asteroseismological analysis and to verify
the pulsation models, it is essential to identify periodicities with
eigen-modes, i.e., obtain their radial order~\n, spherical degree~\l\
and azimuthal order~\m.  For some DB and DO variables, it has been
possible to obtain reasonably sound identifications from a direct
comparison of observed frequencies with model predictions.  However,
for many other stars, including all DA variables, there are either too
few modes, or the mode structure is too variable.

Robinson et al.\ (\cite{robi&a:95}) implemented a method to identify
spherical degree~\l, which combined broad-band measurements of mode
amplitudes at optical and UV wavelengths.  It makes use of the
wavelength dependence of limb darkening, which causes the pulsation
amplitudes to vary with wavelength in a manner that depends on~\l.
Based on its success, we were motivated to implement a variation of
this method which uses high signal-to-noise time-resolved
spectrophotometry at optical wavelengths only.  As will become clear,
our initial data on the star \zzpsc\ showed variations not only in
flux but also in line-of-sight velocity.  The broadened scope, as well
as the increased complexity in modeling the flux variation as a
function of wavelength, made us decide to present the amplitudes and
phases of the modes as a function of wavelength in a separate
paper\footnote{We are far from having exhausted the information in our
data; we will gladly make it available in digital form to anyone
interested in analyzing it further.} (Clemens et al.\
\cite{clemvkw:98}, hereafter Paper~II).

In this paper, we focus on the implications of the line-of-sight
velocity variations.  We first briefly review the basic properties of
\zzpsc\ (\Sec{zzpsc}), and then describe the observation and data
reduction in detail (\Sec{obs}).  In \Sec{ft}, we analyse the
periodicities in the light curve, and in \Sec{vel} we search for
corresponding variations in the line-of-sight velocity curve.  We
discuss the implications of our results in \Sec{disc}, both in a
model-independent way and in the context of the convective driving
models described above.

\section{\zzpsc}\label{sec:zzpsc}

\zzpsc\ (ZZ~Psc; WD~2326+029) was discovered as a variable by Shulov
\& Kopatskaya (\cite{shulk:74}) and studied in some detail by McGraw
and Robinson (\cite{mcgrr:75}).  It is bright ($V\simeq13.0$), at the
cool end of the instability strip ($T=11800\,$K, Bergeron et al.\
\cite{berg&a:95}; cf.\ Paper~II) and shows large-amplitude modulations
(up to a few per cent).  The source remained relatively obscure until
Zuckerman \& Becklin (\cite{zuckb:87}) reported an infrared excess,
which is still unexplained (for recent developments, see Kleinman et
al.\ \cite{klei&a:94}; Koester, Provencal \& Shipman
\cite{koesps:97}).  In 1987, it was first subjected to scrutiny with the
Whole Earth Telescope (WET; Winget et al.\ \cite{wing&a:90}).  Unlike
for the variable DBs and DOs, however, the observations -- while of
excellent quality -- yielded more questions than answers.  Partly,
this is because of the complex and variable frequency spectrum of
\zzpsc: typically, dozens of periodicities are seen, with many
occurring at frequencies that are differences or sums of the
frequencies of other periodicities.

Kleinman et al.\ (\cite{klei&a:98}) discussed an extensive collection
of data sets, obtained over many seasons.  They find that many
periodicities are variable and unstable from year to year, but some do
recur.  By considering the whole ensemble, they could identify with
some certainty which periodicities were associated with real modes and
which with combination frequencies.  They found nineteen real modes,
the majority of which very likely have $\l=1$.  However, due to the
vagaries of the mode characteristics -- especially the enigmatic,
confusing behavior of multiplets, showing different and changing
splitting -- their results are suggestive, but not definitive.

\section{Observations}\label{sec:obs}

Spectra of \zzpsc\ were taken on 1996 November 19 using the Low
Resolution Imaging Spectrometer (Oke et al.\ \cite{oke&a:95}) on the
10-m Keck~II telescope.  From 4:42 till 9:25~{\sc ut}, a continuous
series of 700 12\,s exposures was made.  Only 100 pixels in the
spatial direction (corresponding to $21.5\,$arcsec), binned by a
factor 2, were read out through two amplifiers, leading to a read-out
and re-set time in between exposures of about 12\,s.  The
$600\,{\rm{}line\,mm^{-1}}$ grating was used, covering the range
3450--5950\,\AA\ at $1.2\,{\rm{}\AA\,pix^{-1}}$.  The weather started
photometric, and we used a $8.7\,$arcsec slit to make use of that.
Hence, the seeing of $1.2\,$arcsec determined the wavelength
resolution of $\sim\!7\,$\AA.  In principle, it would have been best
to keep the slit at the parallactic angle in order for refraction not
to influence the wavelength scale, but this would have required
non-standard guiding and we decided to keep the slit at position angle
0\arcdeg\ instead.

At the end of the series, a bank of cloudlets came in, causing the
last 55 spectra, as well as one spectrum some ten minutes earlier, to
have a reduced source and increased sky rate (the moon was up, so
these are easy to identify).  We have not used any of these spectra in
our analysis.

The series of exposures was followed by fifteen 4\,s integrations on
the spectrophotometric flux standard \G191-B2B, using the same setup.
Each of the series was followed by Hg/Kr wavelength calibration and
halogen flat-field frames.

\begin{figure}
\plotone{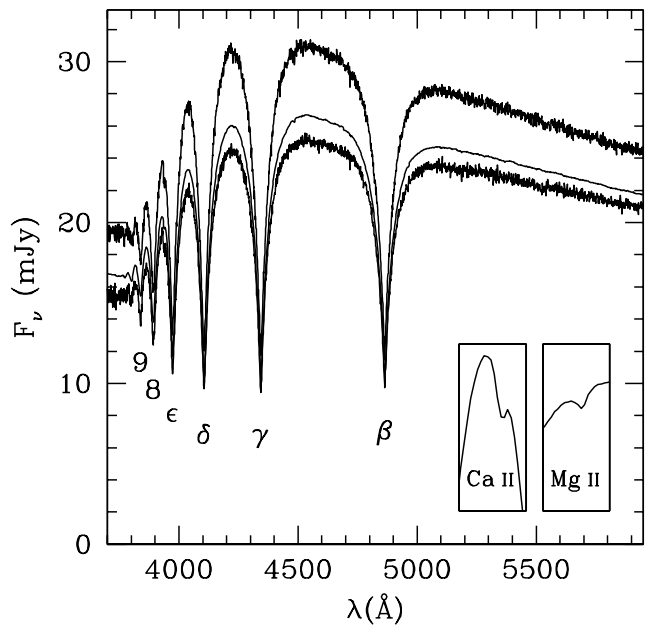}
\caption[]{Average of all spectra and examples of spectra taken at
times of high and low flux (times $-7444$ and $-7080\,$s in
\Fig{lc}b).  Apart from the Balmer lines (marked), weak lines
due to \ion{Ca}{2}~$\lambda3933$ and \ion{Mg}{2}~$\lambda4481$ are
also present (see insets; size 40\,\AA\ by 2\,mJy).  The other
irregularities in the average spectrum are due to pixel-to-pixel
sensitivity variations (see text).\label{fig:spectra}}
\end{figure}

The reduction of all spectra was done as follows.  The frames were
bias-corrected separately for the two amplifiers using the overscan
regions, and then corrected for the gain difference between the two
amplifiers as determined from halogen spectra.  The sky background was
estimated by fitting parabolae for each position in the dispersion
direction, excluding 21 binned pixels ($\sim\!9\arcsec$) centred on
the stellar spectrum.  The latter pixels were used to estimate the
stellar flux using an optimal weighting scheme similar to that of
Horne (\cite{horn:86}).  Note that we did not flat-field the data.
The reason is that we had been unable, due to an instrument failure
shortly after taking the exposures on G191-B2B, to obtain a large
enough number of halogen spectra to construct a flat-field that
matched the signal-to-noise ratio of our observations, especially in
the blue.  Since the flat fields that we did have appeared very
smooth, we decided to forego flat-fielding altogether.  The resulting
spectra look smooth (\Fig{spectra}), except for a few pixels that are
easy to identify.  Note the presence of metals --
\ion{Ca}{2}~$\lambda3933$ and \ion{Mg}{2}~$\lambda4481$ -- uncovered
only recently by Koester et al.\ (\cite{koesps:97}).

Accurate flux calibration was not possible, since the \G191-B2B
spectra were taken while the bank of cloudlets was still passing.  We
selected the three most exposed spectra of the series of fifteen
acquired, as these have comparable count rates and hence may not be
influenced too much, making the relative flux calibration more
reliable\footnote{Perhaps surprisingly, our estimated flux in the
average spectrum compares quite well with the observed $V$ magnitude
of $13.05\pm0.05$ ($22\pm1$\,mJy; McCook \& Sion \cite{mccos:87}).}.
From these three, we derived the instrumental response using the model
fluxes provided by Bohlin, Colina \& Finley (\cite{bohlcf:95}), and
the atmospheric extinction curve of Beland, Boulade \& Davidge
(\cite{belabd:88}).  Typical examples of relatively high and low-flux
spectra, as well as the average spectrum, are shown in
\Fig{spectra}.

\begin{figure}
\plotone{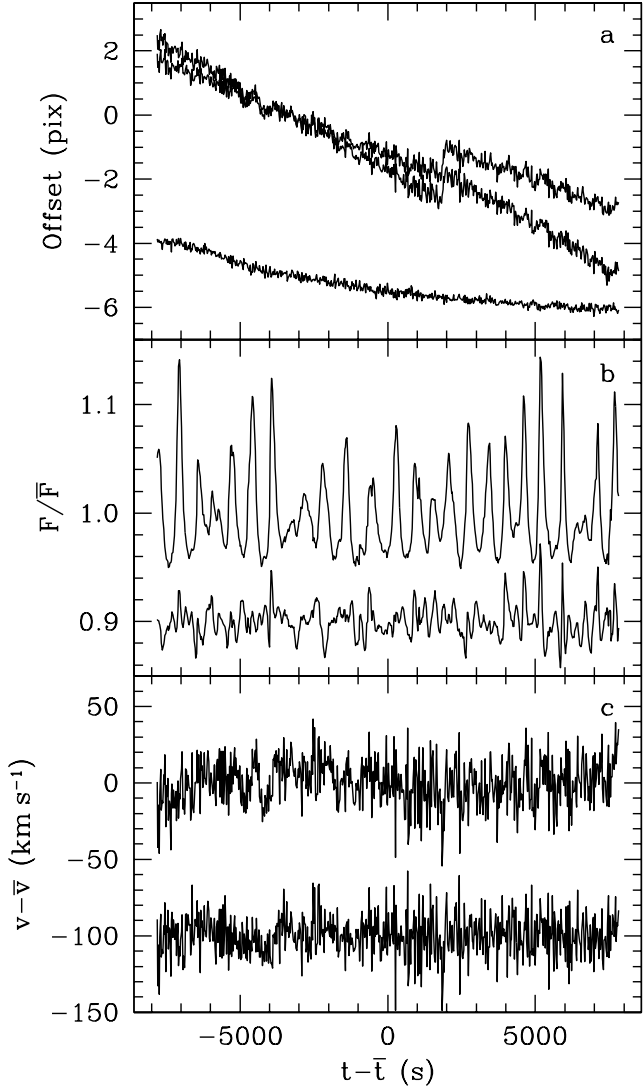}
\caption[]{(a) Variations as a function of time in the pixel position
on the detector of the \ion{O}{1}~$\lambda5577$ sky line (bottom
curve; offset by $-5$ pixels), H$\beta$ (top curve at right) and
H$\epsilon$ (middle curve at right).  The smooth variations are due to
flexure in the spectrograph and differential refraction in the
atmosphere.  (b) Fractional brightness variation (top curve) in the
line-free wavelength region 5200--5500\,\AA, and residual variation
(bottom curve, offset by 0.9) remaining after taking out the eleven
periodic variations with amplitudes larger than 0.55\% and a constant
term (\Fig{ft}a; \Tab{freqs}).  (c) Line-of-sight velocity variations
determined from fits to the H$\beta$, H$\gamma$ and H$\delta$ lines
(top curve), and the residuals (bottom curve, offset by $-100\,\kms$)
remaining after taking out variation at the eleven periods present in
the light curve, as well as a cubic
term.\label{fig:positions}\label{fig:lc}\label{fig:vc}}
\end{figure}

Special care was taken to correct the wavelength scale for
instrumental effects due to flexure and atmospheric refraction.  Both
effects can be seen in the pixel positions on the detector as a
function of time of the \ion{O}{1}~$\lambda5577$ sky line and the
Balmer lines (\Fig{positions}a).  Here, the positions were derived by
cross-correlation of the spectra with the average of those spectra
taken with the slit position angle less than 20\arcdeg\ away from the
parallactic angle.  The break at $t-\bar{t}\simeq2000\,$s in the
Balmer-line positions is due to guiding being stopped and re-started
(to take a guider image).

The flexure is described well by a cubic fit to the sky-line
positions.  After correction for the flexure, as well as for the jump
due to switching the guider off and on, the positions of the Balmer
lines vary linearly with what is expected from refraction
($\sin\alpha\tan\zeta$, where $\alpha$ is the difference between
parallactic and position angle, and $\zeta$ the zenith distance).  The
different slopes indicate that our guiding wavelength was
approximately 7000\,\AA, as expected for the CCD camera used.  We used
this value to calculate the wavelength corrections for the whole
spectrum.

Since our spectra were taken through a wide slit, small movements of
the star will result in random apparent wavelength shifts for
individual spectra.  The effect can be estimated by measuring the
shifts along the slit from the spatial profiles (averaged over some
number of pixels along the dispersion direction).  We find a
root-mean-square variation in these of 0.084 binned pixels,
corresponding to 0\farcs036.  Assuming the variations in the
dispersion direction are similar, one thus expects variations of about
0.2\,\AA, or 14\,\kms\ at H$\gamma$.

Apparent brightness variations can result from varying slit losses due
to seeing-related changes in the width of the stellar images.  From
the spatial profiles, we find 10\% root-mean-square variations around
the average full width at half maximum of 3 binned pixels (1\farcs2).
Using images taken with LRIS for other projects, we estimate that the
slit losses with an $8\farcs7\times9\farcs0$ aperture are about 1\% on
average.  Assuming that the slit losses scale approximately
quadratically with the seeing, one thus expects apparent brightness
variations of the order of 0.2\%.  This is smaller than the Poisson
noise in individual wavelength elements, but dominates in the average
light curve discussed below.  It is also much larger than variations
due to scintillation ($\sim\!0.012\%$; Young \cite{youn:67}).

\begin{figure}
\plotone{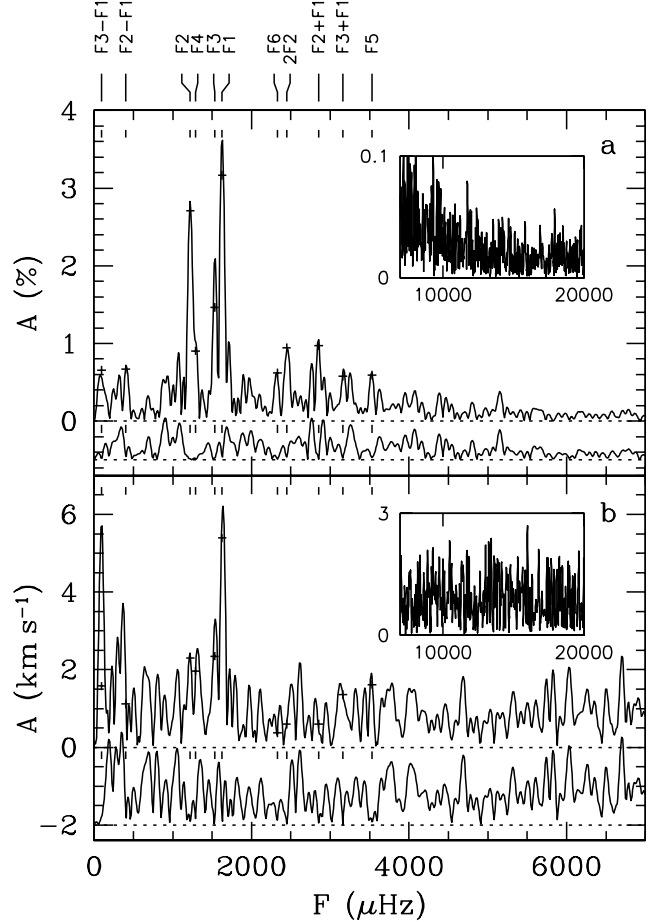}
\caption[]{Fourier transform of the light curve (a) and line-of-sight
velocity curve (b).  The eleven strongest peaks in the Fourier
transform of the light curve are indicated and labeled.  The pluses
indicate the amplitudes derived from fitting eleven cosine waves (plus
a constant for the light curve and a cubic term for the
radial-velocity curve).  The bottom curve in each panel shows the
Fourier transform of the residuals, offset from zero for clarity.  The
insets show the Fourier transform at higher
frequencies.\label{fig:ft}}
\end{figure}

\section{Periodicities}\label{sec:ft}

In \Fig{lc}b, we show the continuum light curve, constructed by
averaging the 5200--5500\,\AA\ region.  The light curve shows the
presence of a number of periodicities, and is similar to light curves
observed before (see \Sec{zzpsc}).  The Fourier transform is shown in
\Fig{ft}a.  Most of the power is at frequencies below 6\,mHz.  An idea
of the background level can be obtained from the 10--20\,mHz region
(see inset in figure), where little evidence for any periodicity is
seen.  In this region, we find a root-mean-square amplitude of about
0.025\%.  The corresponding root-mean-square variation is half that,
and one infers root-mean-square variations in the fluxes of about
$\frac{1}{2}\sqrt{N_{\rm{}data}}\times0.025\%=0.32\%$ (where
$N_{\rm{}data}=644$ is the number of data points), more or less
consistent with the estimate made above based on varying slit losses.
We believe essentially all the signal seen in the main panel of
\Fig{ft}a, which is well above 0.025\%, reflects stellar variability.

We identified the periodicities sequentially.  We started by measuring
the approximate frequency of the highest peak and fitting the data to
a function of the form $C+A\cos(2\pi{}ft'-\phi)$, where
$t'=t-6$:52:08\,{\sc{}ut} is the time relative to the mid-observing
time, solving for pulsation frequency $f$, phase $\phi$ and amplitude
$A$, plus a constant offset $C$.  Next, we identify the highest
remaining peak in the Fourier transform of the residuals and include
another cosine wave in the fit, etc.  For combination frequencies, we
held the frequencies fixed at the corresponding combinations of
parent-mode frequencies.  The decision whether a peak corresponded to
a combination frequency or to a real mode, was generally made
consulting the work of Kleinman (\cite{klei:95}); in all cases, the
amplitude of the combination frequency was the smallest of the peaks
involved.  Continuing as long as the next largest peak had not been
influenced greatly by removing the previous peaks, we found eleven
periodicities with amplitudes larger than 0.55\%.

\begin{table*}
\begin{minipage}{\tabsize}
\caption[]{Strong Pulsation Frequencies in \zzpsc}
\label{tab:freqs}
\begin{tabular}{@{}lrrrrrrrr@{}}
Real mode&\colhead{$P$}&\colhead{$f$}&
\colhead{$A_{\rm L}$}&\colhead{\phnn$\Phi_{\rm L}$}&
\colhead{$A_{\rm V}$}&\colhead{\phnn$\Phi_{\rm V}$}&
\colhead{$R_{\rm V}$}&\colhead{\phnn$\Delta\Phi_{\rm V}$}\\
\colhead{}&\colhead{(s)}&\colhead{($\mu$Hz)}&
\colhead{(\%)}&\colhead{\phnn(\arcdeg)}&
\colhead{(km\,s$^{-1}$)}&\colhead{\phnn(\arcdeg)}&
\colhead{}&\colhead{\phnn(\arcdeg)}\\[1ex]
$\rm\addphn          F1$&  614&\fr{1628.0}{ 9}&\al{3.17}{ 8}&\pl{-168}{ 2}&
    \av{5.4}{ 8}&\pv{-116}{ 8}&\rv{  17}{ 3}&\dv{  52}{ 8}\nl
$\rm\addphn          F2$&  818&\fr{1223.0}{ 9}&\al{2.71}{ 8}&\pl{ 132}{ 2}&
    \av{2.3}{ 8}&\pv{ 176}{19}&\rv{  11}{ 4}&\dv{  44}{19}\nl
$\rm\addphn          F3$&  653&\fr{1531.2}{18}&\al{1.46}{ 8}&\pl{  36}{ 3}&
    \av{2.4}{ 8}&\pv{  59}{19}&\rv{  17}{ 6}&\dv{  23}{19}\nl
$\rm\addphn          F4$&  776&\fr{1288.7}{39}&\al{0.90}{ 8}&\pl{  40}{ 6}&
    \av{2.0}{ 8}&\pv{  50}{22}&\rv{  27}{11}&\dv{  10}{23}\nl
$\rm\addphn          F5$&  283&\fr{3528.8}{49}&\al{0.59}{ 8}&\pl{  55}{ 8}&
    \av{1.6}{ 8}&\pv{ 131}{27}&\rv{  12}{ 6}&\dv{  76}{28}\nl
$\rm\addphn          F6$&  430&\fr{2327.6}{49}&\al{0.62}{ 8}&\pl{ -86}{ 8}&
    \av{0.4}{ 8}&\pv{  -7}{119}&\rv{   4}{ 9}&\dv{  79}{119}\nl[1ex]
Combination&\colhead{$P$}&\colhead{$f$}&
\colhead{$A_{\rm L}$}&\colhead{\phnn$\Phi_{\rm L}$}&
\colhead{$A_{\rm V}$}&\colhead{\phnn$\Phi_{\rm V}$}&
\colhead{$R_{\rm C}$}&\colhead{\phnn$\Delta\Phi_{\rm C}$}\\
frequency&\colhead{(s)}&\colhead{($\mu$Hz)}&
\colhead{(\%)}&\colhead{\phnn(\arcdeg)}&
\colhead{(km\,s$^{-1}$)}&\colhead{\phnn(\arcdeg)}&
\colhead{}&\colhead{\phnn(\arcdeg)}\\[1ex]
$\rm\addphn      F2 +F1$&  351&\fr{  2851.1}{00}&\al{0.97}{ 8}&\pl{ -33}{ 5}&
    \av{0.6}{ 8}&\pv{-137}{72}&\rc{ 5.7}{ 5}&\dc{   3}{ 5}\nl
$\rm\addphn      F2 -F1$& 2469&\fr{   405.0}{00}&\al{0.67}{ 8}&\pl{  34}{ 7}&
    \av{1.1}{ 8}&\pv{ -50}{39}&\rc{ 3.9}{ 5}&\dc{ -26}{ 7}\nl
$\rm\addphn         2F2$&  409&\fr{  2446.1}{00}&\al{0.95}{ 8}&\pl{-131}{ 5}&
    \av{0.6}{ 8}&\pv{ -31}{72}&\rc{12.9}{14}&\dc{ -35}{ 7}\nl
$\rm\addphn      F3 +F1$&  317&\fr{  3159.2}{00}&\al{0.58}{ 8}&\pl{-129}{ 8}&
    \av{1.4}{ 8}&\pv{ -17}{32}&\rc{ 6.2}{10}&\dc{   3}{ 9}\nl
$\rm\addphn      F3 -F1$&10323&\fr{    96.9}{00}&\al{0.66}{ 9}&\pl{ 159}{ 7}&
    \av{1.6}{11}&\pv{-161}{60}&\rc{ 7.1}{10}&\dc{   3}{ 8}\nl
\end{tabular}
\tablecomments{For the real modes,
$\RV=(A_{\rm{}V}/2\pi{}f)/A_{\rm{}L}$ in units of
$10{\rm\,km\,rad^{-1}\,\%^{-1}}$ and
$\DV=\Phi_{\rm{}V}-\Phi_{\rm{}L}$.  For the combination frequencies,
$f=f^i\pm{}f^j$, with for difference frequencies $i$ and $j$ chosen
such that $f$ is positive,
$\RC=A_{\rm{}L}^{i\pm{}j}/n_{ij}A_{\rm{}L}^{i}A_{\rm{}L}^{j}$ (with
$n_{ij}$ the number of possible permutations of $i$ and $j$; unity for
$i=j$, 2 otherwise) and $\DC
=\Phi_{\rm{}L}^{i\pm{}j}-(\Phi_{\rm{}L}^{i}\pm\Phi_{\rm{}L}^{j})$.
The uncertainties in the parameters for the light curve should be seen
as indicating relative uncertainties only, as the light curve is not
described well by the eleven modes fitted and the deviations from the
fit are definitely not distributed normally (see text and \Fig{lc}b).
The estimates given here were obtained by multiplying the
observational errors by such a factor that $\chi^2_{\rm{}red}=1$.  For
the velocity curve, the estimates of the uncertainties should be
reliable.  The uncertainty distributions on amplitude and phase are
not normal, but those on $A\cos\Phi$ and $A\sin\Phi$ are (see
\Sec{velsig}).}
\end{minipage}
\end{table*}

Of the eleven periodicities found, there appear to be six real and
five combination frequencies.  These are listed in \Tab{freqs} and the
corresponding peaks are identified in \Fig{ft}a.  Comparing our
periodicities with those found by Kleinman (\cite{klei:95}), we find
that most periodicities have been seen before.  The exceptions are:
(i) F6: power at 430\,s has only been seen as an apparent harmonic of
a strong 860\,s mode; (ii) F3+F1: power at F1 (614\,s) and F3 (653\,s)
has not been seen simultaneously before, and hence neither has this
combination (although power at 317\,s due to a different combination
has been observed); and (iii) F2$-$F1 and F3$-$F1: at 2469 and
10322\,s, these periods are longer than any generally considered (the
identification of F3$-$F1 should be considered tentative here too).
Absent is power at 400\,s, which has been present in almost all
observations so far.

From the residuals, it is clear that the eleven modes do not account
well for the observed light curve (with our estimate of 0.2\%
root-mean-square uncertainties, we find a formal 
$\chi^2_{\rm{}red}\simeq50$).  We have tried continuing the
identification process (see appendix), but fail to account for all the
variability.  Therefore, we decided to focus on the eleven modes for
which we have full confidence.  We note, however, that because of the
additional power and the resulting non-normal deviations, the formal
errors on the parameters are not very meaningful.  To get an idea of
relative uncertainties, we have scaled up the errors on the data
points such that the reduced $\chi^2$ equals unity.

A puzzling aspect of our results is that the three largest
periodicities appear to have frequencies related to each other by
simple integer ratios: $f_2:f_1\simeq3:4$, $f_2:f_3\simeq4:5$.  Both
are correct to within 0.2\%.  After identifying more modes and
including these in the fit (see appendix), the ratios become more
accurate (to within 0.1\%).  Furthermore, another mode, F8, is
identified for which $f_8:f_1\simeq2:3$, again to within 0.1\%.
Similar close coincidences have not been found in the observations of
Kleinman (\cite{klei:95}) and it may be that they are in some way due
to our relatively short time span, perhaps in combination with the
badness of the fit, especially near the light maxima (see
\Fig{lc}b).  On the other hand, these ratios may represent
unexplored physics underlying the pulsation spectrum.  For the present
time, we will ignore possible complications arising from this.

\section{Velocity variations}\label{sec:vel}

The positions of the Balmer lines on the detector, as derived by the
cross-correlations described in \Sec{obs}, appear to show modulation
on time-scales similar to those shown by the light curve (see
\Fig{positions}a).  This modulation might reflect apparent variations
in the line-of-sight velocity of \zzpsc.  In order to study this in
more detail, we determined Doppler shifts for all the spectra by
fitting the H$\beta$, H$\gamma$ and H$\delta$ lines (details in
\Sec{velsys} below).  The resulting line-of-sight velocity curve is
shown in \Fig{vc}c (upper curve).  The Fourier spectrum, shown in
\Fig{ft}b, shows one strong peak, at a frequency of $1634\pm5\,\mu$Hz,
i.e., consistent with that of the strongest periodicity in the light
curve.

Inspired by the close correspondence between the frequencies of these
signals, we looked for power at frequencies corresponding to the other
ten periodicities in the light curve and found small peaks at some but
not all of them (\Fig{ft}).  We determined their amplitudes and phases
by fitting the velocity curve with a combination of eleven cosine
waves with frequencies fixed at those determined from the light curve
(plus a third-order polynomial to account for slow variations).  The
results are listed in \Tab{freqs} and indicated in \Fig{ft}b; the
residuals are shown in \Fig{vc}c (bottom curve).  Note that unlike for
the light curve, the fit to the line-of-sight velocities is formally
acceptable ($\chi^2_{\rm{}red}\simeq1$) and we expect our error
estimates to be reliable.

Could these line-of-sight velocity variations reflect pulsational
motion? The pulsations are g-modes and the associated motion is
largely horizontal (e.g., Dziembowski \cite{dzie:77}).  The observed
line-of-sight velocity, i.e., the physical velocities projected on the
light of sight and integrated over the stellar disk, will thus be
dominated by motion near the limb.  The observed amplitude will be
smaller than the physical velocity amplitude by an amount depending on
the spherical degree~\l\ and the direction of the pulsation axis
relative to the observer.  Robinson et al.\ (\cite{robikn:82})
estimated the expected line-of-sight velocity amplitudes for typical
pulsation amplitudes and found that they should be a few \kms.  This
is comparable to the amplitudes we find.

If we suppose that the line-of-sight velocity variations reflect
horizontal motion associated with the white dwarf pulsation, it is
very interesting that all five apparently significant detections
($\ge2\sigma$) are for periodicities we identified as real modes
above; no significant detections are found for the combination
frequencies.  Before we turn to the implications, however, we need to
verify that our error estimates are reasonable and the velocity
detections significant; and if so, whether the inferred line-of-sight
velocities reflect intrinsic velocity variations rather than some
systematic effect in our measurement procedure.  We will address these
two points in turn.

\subsection{Significance of the detections}\label{sec:velsig}

Looking at \Fig{ft}b, all peaks but that associated with F1 look
insignificant.  Indeed, they would be formally insignificant if one
did not know the frequency at which to look in advance.  However, the
pulsation frequencies are determined accurately from the light curve.
In the absence of systematic effects (considered separately in
\Sec{velsys}), a rough estimate for the significance of a peak with a
certain amplitude at a frequency specified in advance can be made by
counting the number of other peaks in the velocity spectrum at or
above that amplitude.  For instance, there are 13 peaks with
amplitudes above 2\,\kms\ at frequencies below 7\,mHz (\Fig{ft}b).
With a frequency resolution of $\sim\!64\,\mu$Hz, the total number of
independent frequency elements in this frequency range is 109, and
thus the probability of finding a $\geq2\,\kms$ peak by chance at a
given frequency is roughly 1:8.  One would estimate 1:14 if one counts
peaks up to the Nyquist frequency (21\,mHz; 23 peaks over 2\,\kms\ in
322 [$N_{\rm{}data}/2$] frequency elements).  Thus, F2, F3 and F4, all
of which have amplitudes in excess of $2\,\kms$, are likely real.
There is no other peak as large as F1, hence the probability that it
is a chance coincidence is $<\!1:322$.

\begin{figure}
\plotone{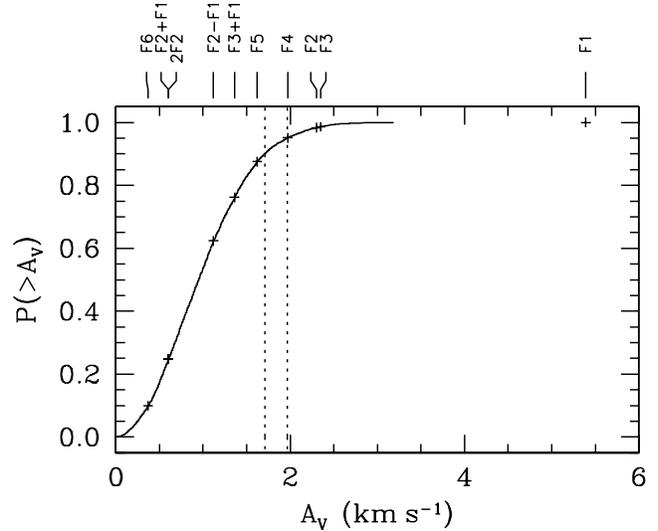}
\caption[]{Cumulative distribution of line-of-sight velocity
amplitudes derived from simulations with scrambled data.  The two
dotted lines indicate the amplitudes at which a measurement at a
frequency specified in advance has a 90 or 95\% probability of being
real (i.e., not due to chance).  The crosses indicate the amplitudes
derived from the fit to the observed line-of-sight velocity curve
(\Fig{vc}).  One sees that the detections of real modes F1, F2, F3 and
F4 are significant ($\simgt\!95$\% confidence), that of F5 is marginal,
and that of F6 is not significant.  No significant signal is measured
for any of the combination frequencies (F3$-$F1 is not shown, because
its probability distribution is different; see text).\label{fig:mc}}
\end{figure}

In order to obtain more quantitative estimates, we ran Monte-Carlo
simulations, in which we fitted scrambled velocity curves in the same
way as we fitted the observations.  The scrambled curves were created
by first taking out the slow third-order variation in the measured
Doppler shifts, and then re-distributing the resulting velocities
randomly over the measurement times.  By counting the number of
artificial data sets for which the fit produces an amplitude larger
than that observed, one can infer the likelihood of a false detection.
The likelihood distributions for all periodicities but F3$-$F1 are
consistent with each other (because of its long period, F3$-$F1 is
covariant with the third-order polynomial, and its likelihood
distribution is wider).  In \Fig{mc}, we show the cumulative
distribution of the amplitudes for all modes but F3$-$F1 derived from
one thousand simulations.  We also indicate the observed amplitudes.
We infer that F1, F2, F3 and F4 are detected at the $>\!99.9$, 98, 98
and 95\% confidence levels, respectively; F5 is marginally detected
($\sim\!88\%$ confidence); and the signals at all other periodicities
are not significant.

As mentioned above, the fit to the line-of-sight velocities is
formally acceptable ($\chi^2_{\rm{}red}\simeq1$).  The errors,
dominated by the star wandering around in the slit (\Sec{obs}), should
be more or less normally distributed, and thus our error estimates
likely reasonable.  Indeed, from the simulations one infers similar
numbers:
$\sigma_A=[\frac{1}{2}\langle{}A^2\rangle]^{1/2}\simeq0.8\,\kms$
(where the factor $\frac{1}{2}$ accounts for the random phase), and
$\sigma_\Phi=(180/\pi)(0.8\,\kms/A_{\rm{}V})\,$deg.  The exception is
F3$-$F1, for which $\sigma_A=1.5\,\kms$, indicating that the
uncertainty for this periodicity is underestimated in the
least-squares fit (\Tab{freqs}).  This reflects the large covariance
with the third-order polynomial.

While the error estimates should be reliable, we caution that the
uncertainty distributions are not expected to be normal.  This is a
consequence of our choice of parameters $A_{\rm{}V}$ and
$\Phi_{\rm{}V}$.  For instance, from \Tab{freqs}, the uncertainty on
$\Phi_{\rm{}V}$ is 28\arcdeg\ for F5.  The $2\sigma$ uncertainty,
however, is far larger than 56\arcdeg: indeed, at 95\% confidence
$\Phi_{\rm{}V}$ can have any value, as the detection of velocity
variations at F5 is significant at $<\!95\%$ confidence.  The
uncertainties on $A_{\rm{}V}\cos\Phi_{\rm{}V}$ and
$A_{\rm{}V}\sin\Phi_{\rm{}V}$ are normally distributed, and one can
use these to determine reliable uncertainty estimates for $A_{\rm{}V}$
and $\Phi_{\rm{}V}$ at different confidence levels.  Note that the
same holds for $A_{\rm{}L}$ and $\Phi_{\rm{}L}$, but the more
significant detections make this less of an issue. 

\subsection{Possible systematic effects}\label{sec:velsys}

Above, we have shown that at frequencies at which the flux of the star
varies strongly, significant variations are also present in the
line-of-sight velocity curve.  The question that remains is whether
these variations reflect the intrinsic motion in the white dwarf's
outer layers expected from the estimates of Robinson et al.\
(\cite{robikn:82}), or whether they might be biased or perhaps even
result from inadequacies in the way we determined the velocities.  For
instance, through changes in continuum slope, the measured Doppler
shift might be influenced by the variations in temperature with
pulsation phase.

For any systematic effect related to the flux variations, the
line-of-sight velocity variations have to be either in phase or in
anti-phase with the light variations, i.e., the phase difference
between velocity and light $\DV=\Phi_{\rm{}V}-\Phi_{\rm{}L}=0$ or
$180\,$deg.  This is not what we observe: for F1, $\DV=52\pm8\,$deg.
Thus, no more than $A_{\rm{}V}\cos\DV=3.3\,\kms$ could be related to
systematic effects of this kind.  The component out of phase with the
light variations has an amplitude of $A_{\rm{}V}\sin\DV=4.2\,\kms$,
which we can be confident is real given the results of the Monte-Carlo
simulations (\Fig{mc}).

While the above shows that at least part of the line-of-sight velocity
signals are intrinsic to the star, the component of the variation that
is in phase or anti-phase with the light variations may still be
influenced.  In order to get a handle on the importance of possible
systematics, we tried different methods to determine the line-of-sight
velocities, looked at differences between the velocities inferred for
different lines, and tried fitting artificial data generated with the
models described in Paper~II.

We found that cross-correlation was very susceptible to systematics.
For instance, the line-of-sight velocity amplitudes derived from the
cross-correlation discussed in \Sec{obs} (\Fig{positions}a) are
systematically different for different Balmer lines, both in phase and
in amplitude.  Most likely, this reflects the influence of the varying
temperature and hence continuum slope, which is not accounted for in
the cross-correlation procedure.  To get unbiased velocities, one
needs to normalize all spectra, but it is not clear how to do this
properly.

We therefore focused on line-profile fitting, of H$\beta$, H$\gamma$
and H$\delta$ (in the wavelength intervals 4622--5102, 4220--4540 and
4040--4200\,\AA, respectively).  To each line, we fitted a sum of a
Lorentzian and a Gaussian profile, with their central wavelengths
forced to be the same, and a continuum represented by either a linear
or a parabolic function.  We tried fitting to $\log{}F$ in addition to
fitting to $F$ directly, the rationale being that since absorption is
a multiplicative rather than an additive process, a sum of components
can represent $\log{}F$ better than $F$.

For all types of fits, the resulting line-of-sight velocity curves
were fitted with the combination of eleven cosine waves.  The results
were systematically different from those inferred from the
cross-correlation velocities, with amplitudes generally being somewhat
smaller, and phase differences between light and velocity being
further away from zero.  The systematic differences between lines were
smaller, but not absent.  No apparent effects are present in the
derived phases for the different periodicities, but the amplitudes
increase systematically from H$\beta$ to H$\gamma$ to H$\delta$.
Also, the zero points (i.e., systemic velocities) derived for the
different lines are very different.  This is probably unrelated,
however, as we find similar differences in systemic velocity for our
flux standard \G191B2B, which has much narrower lines.  Instead, it
likely reflects problems with the wavelength calibration, possibly
related to the way the light of the arc lamp is projected on the slit
(T.~Bida, 1997, private communication).  Such an effect should not
affect the results for the periodic variations in velocity.

This leaves the systematic differences in the amplitudes found for the
individual Balmer lines.  We can discard the possibility that these
are intrinsic to the star based on a simple analytic derivation: while
the intrinsic velocity and light variations are not independent, their
dependency is only a second-order effect and hence very small
(Paper~II).  Most likely, therefore, the observed systematic effects
in amplitude are still related to our way of fitting, perhaps simply
because our choice of Lorentzian plus Gaussian is an inadequate model
for the observed line profiles.

To get a better handle on the problem, we fitted a number of
artificial data sets, produced using the procedures described in
Paper~II, for different settings of temperature amplitude, horizontal
surface velocity amplitude, phase difference between light and
velocity, and spherical degree\footnote{In principle, we could have
tried fitting the models directly to the observations.  We decided not
to do this both because the models do not reproduce the observed
variations all that well, and because it would make verifying our
results much more difficult.}.

For all artificial data sets and fitting methods, the results for the
different lines showed systematic differences of comparable magnitude
to those seen in the real data.  A set with zero input temperature
variation gave the same velocity amplitude and phase for all three
lines, as should be the case, but different systemic velocity, with
H$\gamma$ being close to zero, and H$\beta$ and H$\delta$ being
negative and positive, respectively, by an amount that depended on the
fitting method.  The amount was smaller for fitting $\log{}F$ than for
fitting $F$ directly, as expected ($-13$, $-1$, $+11\,\kms$ for
H$\beta$, H$\gamma$, H$\delta$, respectively).

For data sets with non-zero temperature variations, H$\gamma$ again
reproduced the input values best, in the sense that the systemic value
was still closest to zero, and the velocity phase delay closest to the
input value.  H$\beta$ and H$\delta$ tended to lower and higher
amplitude and to larger and smaller phase delay, respectively, with
the values for both amplitude and phase being more or less symmetric
around those for H$\gamma$.  From a set with no velocity variations
put in, we find that for a temperature variation corresponding to
2.8\% modulation at 5500\,\AA, the apparent velocity shifts in
H$\beta$ and H$\delta$ are about $0.6\,\kms$ in amplitude.  The
systematic differences with H$\gamma$ for sets including velocity
variations are consistent with this.  The amplitude should be compared
with the observed radial velocity amplitude of $\sim\!3\,\kms$ and the
uncertainty of $0.8\,\kms$.

Given the results on the artificial data and taking into account that
the observed velocity signals have low signal-to-noise, we decided to
use the average of the H$\beta$, H$\gamma$ and H$\delta$ velocities,
determined by fitting to $\log{}F$ (with a linear slope for the
continuum).  The resulting line-of-sight velocity curve is the one
discussed above and shown in \Fig{vc}c.  It has uncertainties
dominated by the effects of the star wandering around in the slit.
Given the simulations with the artificial data, we feel reasonably
confident that systematic effects in the fitted amplitudes and phases
will be smaller than the quoted uncertainties derived from the
least-squares fit (as listed in \Tab{freqs}).  For completeness, we
note that the result would be similar if one were to use the
velocities derived from H$\gamma$ only.

\subsection{Velocity summary}\label{sec:velsum}

We conclude that we have detected line-of-sight velocity variations
which are due to intrinsic processes in the white dwarf, and which
have amplitudes comparable with those expected from the integration
over the visible disk of the surface motion due to the pulsations
(Robinson et al.\ \cite{robikn:82}).  We have shown that our estimates
of the uncertainties are robust and that systematic effects, while
problematic, most likely will be within these uncertainties.  Out of
the eleven strongest modes present in the light curve, five out of six
of the real modes show significant modulation, while none of the
combination frequencies does, even though F2+F1 and 2F2 have light
amplitudes similar to those of the weaker real modes F4 and F5.

\section{Discussion}\label{sec:disc}

In this section, we discuss the observational results for the real
modes and the combination frequencies in some detail.  For each group,
we describe the general properties we regard as important, mention
what can be understood from geometrical arguments alone, and discuss
the results in the context of convective-driving models.

\subsection{Real modes}\label{sec:real}

We mentioned already that only the real modes show significant
modulation in the line-of-sight velocities.  As measures of the
relative amplitudes and phases of velocity and light, we use
$\RV=(A_{\rm{}V}/2\pi{}f)/A_{\rm{}L}$ and
$\DV=\Phi_{\rm{}V}-\Phi_{\rm{}L}$ (see \Tab{freqs}).  Here,
$A_{\rm{}V}$ and $A_{\rm{}L}$ are the observed line-of-sight velocity
and fractional light amplitudes, $\Phi_{\rm{}V}$ and $\Phi_{\rm{}L}$
the corresponding phases, and $f$ the frequency of the mode.  The
ratio \RV\ does not depend on azimuthal order~\m\ and aspect (see
Dziembowski \cite{dzie:77}; Paper~II).  The rationale for including
$f$ is that with this definition \RV\ depends only on spherical
degree~\l\ for adiabatic pulsations.  For adiabatic pulsations, one
also has $\DV=90\arcdeg$, i.e., velocity maximum arrives a quarter
cycle after light maximum.  

From \Tab{freqs}, we see that the values of both \RV\ and \DV\ show no
significant variation among the different real modes.  All values of
\DV\ are between 0 and 90\arcdeg, i.e., velocity lags light, but by
less than a quarter cycle.  

The largest value of \RV\ is found for mode~F4.  While the difference
with the values for the other modes is not formally significant, we
note that a larger value is consistent with this mode having $\l=2$
and the others $\l=1$, as inferred in Paper~II from the wavelength
dependences of the pulsation amplitudes.  Compared to an $\l=1$ mode,
the light variations for an $\l=2$ mode are subject to stronger
cancellation, while the surface velocities are actually easier to
observe (see Dziembowski \cite{dzie:77}).

In the interior of a white dwarf, pulsation associated with a real
mode should be largely adiabatic; maximum flux coincides with maximum
compression and is followed a quarter cycle later by matter moving
away at maximum velocity.  This will change as the mode propagates
upward and enters the outer regions, where non-adiabatic effects
become important.  Using the observed relations between light and
line-of-sight velocity variations, we can therefore hope to infer
properties of the surface layers.

In the convective-driving picture (Brickhill \cite{bric:92a};
Goldreich \& Wu \cite{goldw:99a}), the main source of change is the
surface convection zone.  As this region bottles up heat when it is
heated, the flux perturbations at the photosphere of the star are
diminished in magnitude and delayed in phase compared to those
entering the bottom of the convection zone.  In contrast, the
horizontal velocities associated with the pulsation are nearly
independent of depth inside the convection zone, because of the strong
turbulent viscosity (Brickhill \cite{bric:90}; Goldreich \& Wu
\cite{goldw:99b}).  Thus, values of \RV\ should be larger than those
expected for adiabatic pulsations, and values of \DV\ smaller than
90\arcdeg.

The effects are expected to become more prominent with increasing
values of $\omega\tau_C$, where $\omega\equiv2\pi{}f$ is the angular
frequency of the mode and $\tau_C$ is the so-called thermal adjustment
time (Goldreich \& Wu \cite{goldw:99a}; quantity $D$ in Brickhill
\cite{bric:83}), which is a few times the thermal time at the bottom
of the convection zone.  With increasing mode frequency, therefore,
\RV\ should increase and \DV\ should deviate more and more from
90\arcdeg, tending towards~0\arcdeg.

Indeed, the observed values of \DV\ are smaller than 90\arcdeg\ and
those for \RV\ larger by a factor 2--3 than expected if one were to
neglect non-adiabatic effects (e.g., $R_{\rm{}V,ad}\simeq6$ for
mode~F1).  Unfortunately, the observations do not allow one to verify
the predicted trends, but it may be worth noting that, if anything,
\DV\ shows a trend opposite to that predicted, i.e., less of a
deviation from 90\arcdeg\ for shorter periods.  We remind the reader
that the errors in \DV\ are not distributed normally (\Sec{velsig}).

It is difficult to make a more quantitative comparison, as this
requires evaluating the thermal adjustment time $\tau_C$, which
depends on details of the convection.  Ideally, one would like to use
\RV\ and \DV\ to estimate $\tau_C$ and thus constrain the convection
properties.  Our data only allow a rough determination: a 800-s mode
in a star with $\tau_C=300$\,s has $\RV\simeq16$ and
$\DV\simeq50\arcdeg$ (Wu \& Goldreich \cite{wug:98}), similar to what
is observed for mode~F2.

An independent constraint to $\tau_C$ can be inferred from the
longest-period mode that is present.  This is because to excite a
gravity-mode, the convection zone needs to extend deep enough for its
thermal response time to be of order of or longer than the mode
period, i.e., $\omega\tau_C>1$ (Brickhill \cite{bric:83}; Goldreich \&
Wu \cite{goldw:99a}).  The longest-period mode is F2 at 818\,s (or F7
at 1106\,s; see Appendix), and thus $\tau_C\simgt130\,$s
($\simgt\!180\,$s), consistent with the above value.  Consistent
estimates of $\tau_C$ are also obtained using combination frequencies
(see below).

\subsection{Combination frequencies}\label{sec:comb}

The lack of line-of-sight velocity signals at the frequencies which we
associate with combinations of real modes, confirms the notion that
combination frequencies do not reflect physical pulsation, but rather
mixing of the parent-mode signals by a non-linear transformation
occurring in the outer layers of the white dwarf.  

We quantify the combination frequencies using an amplitude ratio
$\RC=A_{\rm{}L}^{i\pm{}j}/n_{ij}A_{\rm{}L}^iA_{\rm{}L}^j$ and a phase
difference
$\DC=\Phi_{\rm{}L}^{i\pm{}j}-(\Phi_{\rm{}L}^i\pm\Phi_{\rm{}L}^j)$.
Here, $A_{\rm{}L}^{i\pm{}j}$ and $\Phi_{\rm{}L}^{i\pm{}j}$ are the
observed amplitude and phase of a combination at frequency
$f^i\pm{}f^j$ of real modes $i$ and $j$ (with frequencies $f^i$ and
$f^j$, amplitudes $A_{\rm{}L}^i$ and $A_{\rm{}L}^j$, and phases
$\Phi_{\rm{}L}^i$ and $\Phi_{\rm{}L}^j$); $n_{ij}$ is the number of
possible permutations: $n_{ij}=1$ for $i=j$ and $n_{ij}=2$ for
$i\neq{}j$; $i$ and $j$ are chosen such that difference frequencies
are positive.  We include $n_{ij}$ to ensure that the harmonic of a
single mode and a combination of two modes with almost identical
properties would lead to the same \RC\ (cf.~Brickhill
\cite{bric:92b}).  Similar definitions are used for the 3-mode
combinations (see \Tab{freqs-app}); below, we focus on combinations of
two modes.

The observed values of \RC\ and \DC\ (Tables~\ref{tab:freqs}
and~\ref{tab:freqs-app}) vary significantly from mode to mode,
although part of the variation is due to only a few outliers
(especially 2F4, which we will discuss in more detail below).  There
appears to be some correlation between the value of \RC\ and the
amplitudes of the parent modes, with \RC\ increasing for
decreasing~$A_{\rm{}L}^{i,j}$.  The phase differences \DC\ cluster
around zero (within $\sim\!45\arcdeg$); this reflects the sharp maxima
and shallow minima in the light curve.

If the combination frequencies indeed result from nonlinear mixing in
the upper layers of the star, then independent of a specific theory
one expects a combination frequency to have at any point on the
surface a flux amplitude proportional to the product of its
parent-mode amplitudes at that point.  The surface distributions of
the parent modes are described by spherical harmonics.  Therefore, the
surface distribution of the combination frequency will be proportional
to the product of spherical harmonics, which can in turn be described
by a linear superposition of spherical harmonics with different~\l\
(see, e.g., Abramowitz \& Stegun \cite{abras:72}; we adopt their
notation).

In the following discussion, we will assume all parent modes have only
$\m=0$ components.  We consider the more general case later.  First,
we study the first harmonic of a single parent mode.  The first
harmonic of a parent mode with $(\l,\m)=(1,0)$ has a surface
distribution described by $Y_1^0 Y_1^0$, which can be decomposed into
a sum of $Y_0^0$ and $Y_2^0$ distributions.  The first harmonic of a
parent mode with $(\l,\m)=(2,0)$ has a surface distribution that is
the sum of three components, $Y_0^0$, $Y_2^0$ and $Y_4^0$.  The
numerical factors in front of the $Y_0^0$ terms are the same for both
decompositions.  When integrated over the visible hemisphere, the
(2,0) parent mode will suffer stronger cancellation than the (1,0)
parent mode.  On the other hand, their harmonics will typically be
dominated by $Y_0^0$ components that do not suffer from cancellation.
Therefore, \RC\ will be higher for the harmonic of the (2,0) mode.
This conclusion is independent of the the angle between the
line-of-sight and the pulsation axis.

In \Tab{freqs-app}, the value of \RC\ for 2F4 is much larger than,
e.g., that for 2F1 and 2F2.  While we should caution that 2F4 could be
blended with other combinations, this is exactly what one would expect
from the example given above, given our identifications of F1 and F2
with $\l=1$ and F4 with $\l=2$ (Paper~II).  Note that one would not
expect, e.g., F4+F1 to have a value of \RC\ larger than those for
combinations of two $\l=1$ modes, such as F2+F1, F3+F1 and F5+F1 (as
observed; \Tab{freqs-app}).  This is because the product $Y_2^0Y_1^0$
has no $Y_0^0$ component and therefore suffers cancellation.

Looking in detail at \RC\ for all combinations involving F1 and
another $\l=1$ mode, e.g., 2F1, F2+F1, we find that the values are not
consistent with being the same.  There appears to be a trend of
increasing \RC\ with decreasing parent-mode amplitude.  A similar
trend appears to be present for the 3-mode combination frequencies.
This is not expected, as generally the efficiency of mixing should be
relatively independent of the amplitudes of the input signals when
these signals are weak.

We should caution that above we have been neglecting the possible
effect of different values of~\m.  For instance, the harmonic of a
$\l=1$, $\m\neq0$ parent mode is described by a $Y_2^{2m}$
distribution only; it would have a low value of \RC.  In our short
observations, we cannot resolve the rotation-induced frequency
splitting among different~\m\ components that has been seen previously
for some of the modes (Kleinman et al.\ \cite{klei&a:98}).  Thus, the
power at both the harmonic and its parent mode may result from a
number of combinations, which makes our conclusions uncertain.  They
can likely be verified, however, using existing, much longer time
series from WET.

We now turn from these general considerations, valid for any local
nonlinear mixing scheme, to a comparison of our data with numerical
results and analytic calculations based on the convective-driving
mechanism.  As mentioned, in this picture flux perturbations appearing
at the photosphere are diminished and delayed relative to those that
enter the convection zone, by an amount that depends on the depth of
the convection zone (and thus $\tau_C$).  The depth of the convection
zone, however, depends sensitively on the photospheric temperature.
The latter varies due to the pulsations, and, therefore, the depth of
the convection will vary as well.  The associated variation in
reduction and delay leads to a distortion of the pulsation signal at
the photosphere, which translates into the power seen at the
combination frequencies (Brickhill \cite{bric:92b}; Wu \cite{wu:97}).

Brickhill (\cite{bric:92b}) found that \RC\ will be $\simgt\!6$ for
the first harmonic of an $(\l,\m)=(1,0)$ mode and $\simgt\!16$ for
that of a (2,0) mode, the exact values depending on the frequencies,
$\tau_C$ and the angle between the line of sight and the pulsation
axis.  He found sum frequencies share similar values of \RC\ with the
harmonics\footnote{Brickhill (\cite{bric:92b}) discusses but does not
correct for the number of permutations $n_{ij}$.}, while difference
frequencies from the same pair of parent modes have lower values.
These results were confirmed analytically by Wu (\cite{wu:97}), who
found that for combinations with relatively high frequency
($(\omega^{i\pm{}j}\tau_C\gg1$), \RC\ should be approximately constant
and \DC\ approximately zero, while towards lower frequencies
($\omega^{i\pm{}j}\tau_C\simeq1$), \RC\ should be lower and \DC\
should shift towards~$-90\arcdeg$ .

The observed values of \RC\ and \DC\ compare, on average, favourably
to the expectations. The typical values of \RC\ are reproduced with
$\tau_C\simeq200\,$s~(Wu \cite{wu:97}), which is consistent with the
values of $\tau_C$ inferred above from the longest-period overstable
mode and the velocity-light comparison (\Sec{real}).

Little evidence is found for the expected trends of either \RC\ or
\DC\ with frequency of the combinations.  In part, this might be due
to our inability to resolve components of different~\m.  Again, it
would be good to verify these conclusions with longer time series,
such as those already available from WET.

As mentioned, a trend of \RC\ with parent-mode amplitude is not
expected in any simple non-linear mixing theory (e.g.,
Wu~\cite{wu:97}).  Nevertheless, the numerical calculations presented
by Brickhill (\cite{bric:92b}) show an amplitude dependence of \RC\
for 3-mode combinations that is similar to that observed (and with
similar numerical values).  At present, we do not understand why this
dependence arises in the simulations; unfortunately, Brickhill does
not expand on it.  

\section{Conclusions}\label{sec:conc}

We have applied the tool of time-resolved spectrophotometry to the
study of pulsating white dwarfs.  In this paper, we showed that at our
very high signal-to-noise ratios we can detect the small line-of-sight
velocity variations associated with the g-mode pulsations, at
amplitudes similar to those expected (Robinson et al.\
\cite{robikn:82}).  Significant velocity variation is seen for five
out of the six modes we believed were real.  For all five modes, the
phase difference \DV\ between velocity and light is between 0 and
90\arcdeg, i.e., velocity maximum occurs later than flux maximum, but
by less than the quarter cycle expected for adiabatic pulsations.  The
amplitude ratios \RV\ are similar for the five modes.  The largest
value occurs for mode~F4 (at 776\,s), which is consistent with mode F4
having $\l=2$ and the others $\l=1$, as was inferred from the
wavelength dependence of the pulsation amplitudes in Paper~II.

No significant line-of-sight velocity variation is seen for any of the
five combination frequencies, even though two of these have light
amplitudes comparable to that of the fourth strongest real mode.  This
provides independent confirmation that the combination frequencies do
not reflect physical pulsation.  Rather, they likely result from
mixing of mode power by a non-linear transformation in the outer
layers of the star.  The phase differences \DC\ between combination
frequencies and parent modes are all close to zero, reflecting the
sharp maxima and shallow minima in the light curve.  The amplitude
ratios \RC\ show significant differences, but are generally within a
factor two of each other.  The exception is the four times larger
value for the harmonic of mode~F4.  This difference is again
consistent with the differences in~\l\ between mode F4 and the others.

Models for the interaction between pulsation and convection can
broadly reproduce the observed values of \RV\ and \DV\ for the real
modes, as well as those of \RC\ and \DC\ for the combination
frequencies (Brickhill \cite{bric:92b}; Wu \cite{wu:97}).  The models
are internally consistent, in that the inferred thermal properties of
the convection zone are such that modes at the observed periods are
expected to be overstable.  In detail, there are some problems, with
\DV\ showing a trend with frequency opposite of that expected, and the
mixing strength, reflected in \RC, not showing the expected
correlation with frequency, but rather a possible dependency on
parent-mode amplitude.  Both these need confirmation; for the former,
further time-resolved spectroscopy is needed, but for the latter,
excellent data already exist.  We also hope that our results will lead
to further theoretical efforts.

In summary, time-resolved spectroscopy offers insight in the white
dwarf interiors at three levels.  By measuring the effects of
temperature variation at different wavelengths, the temperature
structure of the atmosphere can be calibrated.  Using the velocity
changes and the combination frequencies, one can constrain the
convection zone below the photosphere.  And finally, with the
determination of the spherical degree and hence the more secure
identifications of periodicities with eigen-modes, one can use
asteroseismology to probe the interior with greater confidence.

\acknowledgements We gratefully acknowledge help of Wayne Wack with
the observations, useful discussions with Tom Bida about intricacies
of LRIS and with Peter Goldreich, Scot Kleinman and Rob Robinson about
white-dwarf pulsations, as well as sceptical and therefore very useful
remarks by an anonymous referee.  M.H.v.K.\ acknowledges a NASA Hubble
Fellowship while at Caltech, and a fellowship of the Royal Netherlands
Academy of Arts and Sciences at Utrecht.  This research has made use
of the Simbad database, operated at CDS, Strasbourg, France.  The
observations were obtained at the W.~M.~Keck Observatory, which is
operated by the California Association for Research in Astronomy, a
scientific partnership among the California Institute of Technology,
the University of California, and the National Aeronautics and Space
Administration.  It was made possible by the generous financial
support of the W.~M.~Keck foundation.  The reduction was done within
the environment of the Munich Image Data Analysis System, which is
developed and maintained by the European Southern Observatory.

\appendix
\section{Further Periodicities}

\begin{figure}
\plotone{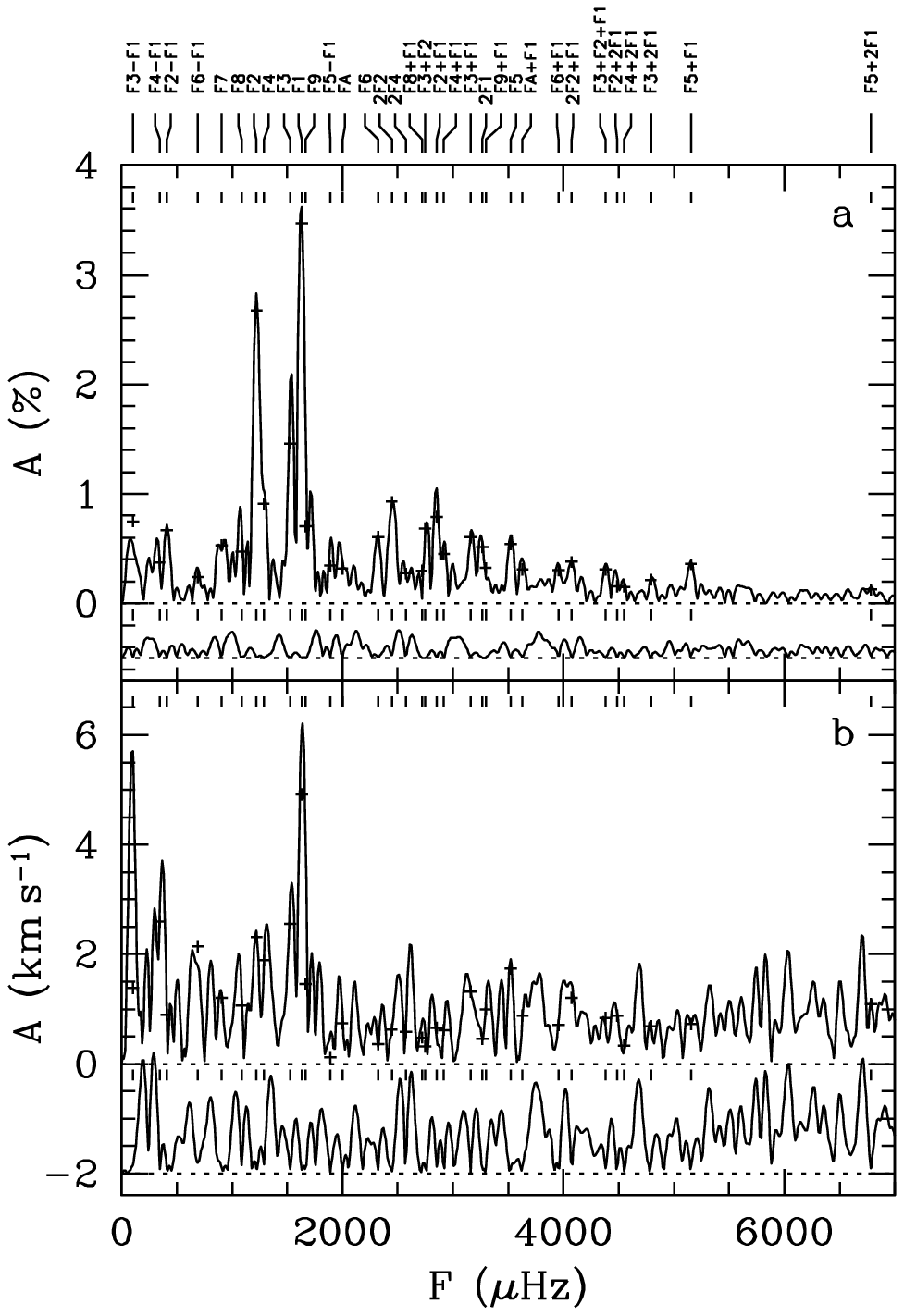}
\caption[]{Fourier transform of the light curve (a) and line-of-sight
velocity curve (b), with 33 identified peaks indicated and labeled.
The pluses indicate the amplitudes derived from fitting cosine waves
(plus a constant for the light curve and a cubic polynomial for the
line-of-sight velocity curve).  The bottom curve in each panel shows
the Fourier transform of the residuals, offset from zero for
clarity.\label{fig:ft-app}}
\end{figure}

The light curve we obtained is not described well by variations at the
eleven frequencies we identified in \Sec{ft}.  In particular,
there are a quite a few more peaks present in the Fourier transform at
combination frequencies.  We have continued the identification process
until no peaks with amplitudes larger than 0.3\% remained.  This gave
us 29 frequencies\footnote{The fitted amplitude is not always larger
than 0.3\% for combination frequencies, as for these the frequency is
fixed at the sum or difference of the frequencies of the parent modes,
which may not lead to a complete removal of the peak in the Fourier
transform.}.  At this point, the Fourier spectrum of the residuals
became too noisy to identify with any confidence more peaks in the
main region of power (1--4\,mHz).  In the cleaner region at higher
frequencies, however, there were still four peaks coincident with
frequencies expected for 3-mode combinations.  Since these might be of
particular interest for modeling the pulsations, we included these in
a final fit.  The frequencies, amplitudes and phases for all 33
periodicities are listed in \Tab{freqs-app}, and indicated in
\Fig{ft-app}a.  We also fitted the velocities to this set of
33 periodicities.  The results are indicated in
\Fig{ft-app}b; calculated values of \RV\ and \DV\ for the
real modes are listed in \Tab{freqs-app}.

We conclude with a few remarks.  First, we note that possible
significant periodicity is also present for 2F3
($A_{\rm{}L}=0.21\pm0.05\%$, $\Phi_{\rm{}L}=-3\pm14\arcdeg$,
$\RC=9.7\pm2.4$, $\DC=-83\pm14\arcdeg$) and F5+F2 ($0.12\pm0.05\%$,
$-117\pm26\arcdeg$, $4.0\pm1.9$, $62\pm26\arcdeg$), while no
significant signal is seen at 2F5 ($0.05\pm0.05\%$, $\RC=17\pm17$).
One might have expected power at a number of combination frequencies
not present in \Tab{freqs-app}, such as F3$-$F2 and F7$\pm$F1.  These
may indeed be present, but their frequencies are so close to other
combinations that it is impossible to include them in the fit.  Which
of the possible combinations one uses is a matter of choice.  We have
been biased by theory in preferring those for which the parent modes
have larger amplitudes, or, if similar, those for which the fitted
value of \DC\ was closest to~0.

Second, the periodicity F9, which is very close to F1 in period, is
rather uncertain.  The modulation could be fitted equally well with a
variable amplitude for~F1.  If F9 is real, it should have $\l\neq1$
given its proximity to F1, which has $\l=1$.  If it has $\l=2$ and a
substantial $\m=0$ component, one might expect a large harmonic, like
for~F4.  This harmonic might in part account for the peak now
described by F9+F1; if we fit 2F9 instead (we cannot fit both
jointly), we find $A_{\rm{}L}=0.26\pm0.05\%$,
$\Phi_{\rm{}L}=109\pm12\arcdeg$, which corresponds to $\RC=36\pm17$,
similar to that of 2F4, but $\DC=-160\pm20\arcdeg$, much further
removed from zero than for any other combination frequency.

Finally, while for most combination frequencies we do not find
significant velocity variation, there are two exceptions: F4$-$F1 at
2900\,s and F6$-$F1 at 1447\,s.  These two periodicities appear
harmonically related to each other, and we believe the associated
velocity signals reflect either instrumental or atmospheric effects
rather than physical variations.  We briefly wondered about possible
orbital modulation, but observations at higher resolution did not
confirm these periodicities.

\begin{table*}
\begin{minipage}{\apptabsize}
\caption[]{Pulsation Frequencies in \zzpsc}
\label{tab:freqs-app}
\begin{tabular}{@{}lrrrrrrrr@{}}
Real mode&\colhead{$P$}&\colhead{$f$}&
\colhead{$A_{\rm L}$}&\colhead{\phnn$\Phi_{\rm L}$}&
\colhead{$A_{\rm V}$}&\colhead{\phnn$\Phi_{\rm V}$}&
\colhead{$R_{\rm V}$}&\colhead{\phnn$\Delta\Phi_{\rm V}$}\\
\colhead{}&\colhead{(s)}&\colhead{($\mu$Hz)}&
\colhead{(\%)}&\colhead{\phnn(\arcdeg)}&
\colhead{(km\,s$^{-1}$)}&\colhead{\phnn(\arcdeg)}&
\colhead{}&\colhead{\phnn(\arcdeg)}\\[1ex]
$\rm\addphn          F1$&  613&\fr{1631.5}{ 9}&\al{3.47}{14}&\pl{-174}{ 2}&
    \av{4.9}{10}&\pv{-122}{12}&\rv{  14}{ 3}&\dv{  52}{12}\nl
$\rm\addphn          F2$&  818&\fr{1222.4}{ 6}&\al{2.67}{ 5}&\pl{ 132}{ 1}&
    \av{2.3}{ 8}&\pv{ 177}{19}&\rv{  11}{ 4}&\dv{  45}{19}\nl
$\rm\addphn          F3$&  655&\fr{1527.5}{11}&\al{1.46}{ 5}&\pl{  40}{ 2}&
    \av{2.6}{ 8}&\pv{  61}{18}&\rv{  19}{ 6}&\dv{  21}{18}\nl
$\rm\addphn          F4$&  777&\fr{1286.7}{17}&\al{0.91}{ 5}&\pl{  44}{ 4}&
    \av{1.9}{ 8}&\pv{  53}{23}&\rv{  26}{11}&\dv{   9}{23}\nl
$\rm\addphn          F5$&  284&\fr{3521.5}{25}&\al{0.54}{ 5}&\pl{  50}{ 5}&
    \av{1.7}{ 8}&\pv{ 132}{25}&\rv{  14}{ 7}&\dv{  82}{25}\nl
$\rm\addphn          F6$&  431&\fr{2322.5}{26}&\al{0.61}{ 5}&\pl{ -85}{ 5}&
    \av{0.4}{ 8}&\pv{ -34}{121}&\rv{   4}{ 9}&\dv{  51}{121}\nl
$\rm\addphn          F7$& 1106&\fr{ 903.9}{34}&\al{0.53}{ 5}&\pl{-125}{ 5}&
    \av{1.2}{ 8}&\pv{ -23}{37}&\rv{  40}{27}&\dv{ 102}{37}\nl
$\rm\addphn          F8$&  920&\fr{1086.7}{39}&\al{0.47}{ 5}&\pl{ 141}{ 6}&
    \av{1.1}{ 8}&\pv{  99}{40}&\rv{  34}{25}&\dv{ -42}{40}\nl
$\rm\addphn          F9$&  601&\fr{1664.9}{60}&\al{0.71}{14}&\pl{ -43}{ 8}&
    \av{1.5}{10}&\pv{ -63}{38}&\rv{  20}{14}&\dv{ -20}{39}\nl
$\rm\addphn          FA$&  500&\fr{1998.4}{42}&\al{0.32}{ 5}&\pl{  77}{ 9}&
    \av{0.7}{ 8}&\pv{  82}{59}&\rv{  17}{20}&\dv{   5}{60}\nl[1ex]
2-mode&\colhead{$P$}&\colhead{$f$}&
\colhead{$A_{\rm L}$}&\colhead{\phnn$\Phi_{\rm L}$}&
\colhead{$A_{\rm V}$}&\colhead{\phnn$\Phi_{\rm V}$}&
\colhead{$R_{\rm C}$}&\colhead{\phnn$\Delta\Phi_{\rm C}$}\\
combination&\colhead{(s)}&\colhead{($\mu$Hz)}&
\colhead{(\%)}&\colhead{\phnn(\arcdeg)}&
\colhead{(km\,s$^{-1}$)}&\colhead{\phnn(\arcdeg)}&
\colhead{}&\colhead{\phnn(\arcdeg)}\\[1ex]
$\rm\addphn         2F1$&  306&\fr{  3263.0}{00}&\al{0.52}{ 8}&\pl{ -10}{ 9}&
    \av{0.5}{10}&\pv{ 122}{123}&\rc{ 4.3}{ 7}&\dc{ -22}{10}\nl
$\rm\addphn      F2 +F1$&  350&\fr{  2853.9}{00}&\al{0.79}{ 5}&\pl{ -35}{ 4}&
    \av{0.7}{ 8}&\pv{-143}{69}&\rc{ 4.3}{ 3}&\dc{   7}{ 5}\nl
$\rm\addphn      F2 -F1$& 2444&\fr{   409.2}{00}&\al{0.67}{ 5}&\pl{  29}{ 4}&
    \av{0.9}{ 8}&\pv{ -42}{49}&\rc{ 3.6}{ 3}&\dc{ -25}{ 5}\nl
$\rm\addphn         2F2$&  409&\fr{  2444.7}{00}&\al{0.93}{ 5}&\pl{-132}{ 3}&
    \av{0.6}{ 8}&\pv{ -39}{69}&\rc{13.0}{ 9}&\dc{ -37}{ 4}\nl
$\rm\addphn      F3 +F1$&  317&\fr{  3159.0}{00}&\al{0.61}{ 5}&\pl{-117}{ 5}&
    \av{1.3}{ 8}&\pv{  -9}{34}&\rc{ 6.0}{ 6}&\dc{  17}{ 6}\nl
$\rm\addphn      F3 -F1$& 9612&\fr{   104.0}{00}&\al{0.75}{ 5}&\pl{ 158}{ 4}&
    \av{1.4}{13}&\pv{-128}{46}&\rc{ 7.4}{ 7}&\dc{  11}{ 5}\nl
$\rm\addphn      F3 +F2$&  364&\fr{  2749.9}{00}&\al{0.69}{ 8}&\pl{ 147}{ 6}&
    \av{0.3}{11}&\pv{ 165}{191}&\rc{ 8.8}{11}&\dc{ -26}{ 6}\nl
$\rm\addphn      F4 +F1$&  343&\fr{  2918.2}{00}&\al{0.45}{ 5}&\pl{-155}{ 7}&
    \av{0.6}{ 8}&\pv{ -39}{72}&\rc{ 7.1}{ 9}&\dc{ -25}{ 8}\nl
$\rm\addphn      F4 -F1$& 2900&\fr{   344.9}{00}&\al{0.38}{ 5}&\pl{-163}{ 8}&
    \av{2.6}{ 8}&\pv{ -38}{17}&\rc{ 6.0}{ 9}&\dc{  54}{ 9}\nl
$\rm\addphn         2F4$&  389&\fr{  2573.3}{00}&\al{0.27}{ 5}&\pl{  91}{10}&
    \av{0.6}{ 8}&\pv{  97}{76}&\rc{33.2}{71}&\dc{   4}{13}\nl
$\rm\addphn      F5 +F1$&  194&\fr{  5153.1}{00}&\al{0.36}{ 5}&\pl{-127}{ 8}&
    \av{0.7}{ 8}&\pv{  44}{59}&\rc{ 9.5}{16}&\dc{  -3}{10}\nl
$\rm\addphn      F5 -F1$&  529&\fr{  1890.0}{00}&\al{0.35}{ 5}&\pl{-129}{ 8}&
    \av{0.1}{ 8}&\pv{-174}{365}&\rc{ 9.2}{16}&\dc{   7}{10}\nl
$\rm\addphn      F6 +F1$&  253&\fr{  3954.0}{00}&\al{0.30}{ 5}&\pl{ 152}{ 9}&
    \av{0.7}{ 8}&\pv{ -52}{61}&\rc{ 7.2}{13}&\dc{  50}{11}\nl
$\rm\addphn      F6 -F1$& 1447&\fr{   690.9}{00}&\al{0.24}{ 5}&\pl{  55}{12}&
    \av{2.1}{ 8}&\pv{-130}{20}&\rc{ 5.7}{13}&\dc{ -34}{13}\nl
$\rm\addphn      F8 +F1$&  368&\fr{  2718.2}{00}&\al{0.30}{ 8}&\pl{ -32}{13}&
    \av{0.5}{10}&\pv{ 171}{124}&\rc{ 9.1}{27}&\dc{   1}{15}\nl
$\rm\addphn      F9 +F1$&  303&\fr{  3296.4}{00}&\al{0.32}{ 8}&\pl{ 114}{13}&
    \av{1.0}{10}&\pv{ -88}{57}&\rc{ 6.6}{21}&\dc{ -30}{15}\nl
$\rm\addphn F\nsiz A+F1$&  275&\fr{  3629.9}{00}&\al{0.31}{ 5}&\pl{ -87}{ 9}&
    \av{0.9}{ 8}&\pv{ 133}{50}&\rc{14.0}{32}&\dc{   9}{13}\nl[1ex]
3-mode&\colhead{$P$}&\colhead{$f$}&
\colhead{$A_{\rm L}$}&\colhead{\phnn$\Phi_{\rm L}$}&
\colhead{$A_{\rm V}$}&\colhead{\phnn$\Phi_{\rm V}$}&
\colhead{$R_{\rm C}$}&\colhead{\phnn$\Delta\Phi_{\rm C}$}\\
combination&\colhead{(s)}&\colhead{($\mu$Hz)}&
\colhead{(\%)}&\colhead{\phnn(\arcdeg)}&
\colhead{(km\,s$^{-1}$)}&\colhead{\phnn(\arcdeg)}&
\colhead{}&\colhead{\phnn(\arcdeg)}\\[1ex]
$\rm\addphn     2F2 +F1$&  245&\fr{  4076.3}{00}&\al{0.38}{ 5}&\pl{  55}{ 8}&
    \av{1.2}{ 8}&\pv{-129}{36}&\rd{  51}{ 7}&\dd{ -36}{ 8}\nl
$\rm\addphn  F3 +F2 +F1$&  228&\fr{  4381.4}{00}&\al{0.31}{ 5}&\pl{ -11}{ 9}&
    \av{0.8}{ 8}&\pv{-163}{53}&\rd{  39}{ 7}&\dd{  -9}{10}\nl
$\rm\addphn      F2+2F1$&  223&\fr{  4485.4}{00}&\al{0.16}{ 5}&\pl{ 136}{18}&
    \av{0.9}{ 8}&\pv{-124}{50}&\rd{  16}{ 5}&\dd{  -9}{19}\nl
$\rm\addphn      F3+2F1$&  209&\fr{  4790.5}{00}&\al{0.22}{ 5}&\pl{  69}{13}&
    \av{0.7}{ 8}&\pv{ 179}{63}&\rd{  41}{10}&\dd{  16}{14}\nl
$\rm\addphn      F4+2F1$&  220&\fr{  4549.7}{00}&\al{0.15}{ 5}&\pl{  18}{19}&
    \av{0.3}{ 8}&\pv{   3}{127}&\rd{  46}{16}&\dd{ -38}{20}\nl
$\rm\addphn      F5+2F1$&  147&\fr{  6784.6}{00}&\al{0.13}{ 5}&\pl{  78}{22}&
    \av{1.1}{ 8}&\pv{ -76}{40}&\rd{  64}{26}&\dd{  15}{23}\nl
\end{tabular}
\tablecomments{For the real modes,
$\RV=(A_{\rm{}V}/2\pi{}f)/A_{\rm{}L}$ in units of
$10{\rm\,km\,rad^{-1}\,\%^{-1}}$ and
$\DV=\Phi_{\rm{}V}-\Phi_{\rm{}L}$.  For the 2-mode combination
frequencies, $f=f^i\pm{}f^j$, with for difference frequencies $i$ and
$j$ chosen such that $f$ is positive,
$\RC=A_{\rm{}L}^{i\pm{}j}/n_{ij}A_{\rm{}L}^{i}A_{\rm{}L}^{j}$, with
$n_{ij}$ the number of possible permutations, and
$\DC=\Phi_{\rm{}L}^{i\pm{}j}-
(\Phi_{\rm{}L}^{i}\pm\Phi_{\rm{}L}^{j})$.  Similarly, for 3-mode
combination frequencies, $f=f^i+f^j+f^k$, $\RC=A_{\rm{}L}^{i+j+k}/
n_{ijk}A_{\rm{}L}^{i}A_{\rm{}L}^{j}A_{\rm{}L}^{k}$ and
$\DC=\Phi_{\rm{}L}^{i+j+k}-
(\Phi_{\rm{}L}^{i}+\Phi_{\rm{}L}^{j}+\Phi_{\rm{}L}^{k})$.  The
uncertainties in the parameters for the light curve should be seen as
indicating relative uncertainties only, as the light curve is not
described well by the modes fitted and the deviations from the fit are
definitely not distributed normally.  The estimates given here were
obtained by multiplying the observational errors by such a factor that
$\chi^2_{\rm{}red}=1$.  For the velocity curve, the estimates of the
uncertainties should be reliable.  The uncertainty distributions on
amplitude and phase are not normal, but those on $A\cos\Phi$ and
$A\sin\Phi$ are (see \Sec{velsig}).}
\end{minipage}
\end{table*}
\end{document}